\documentclass[fleqn,10pt]{wlscirep}

\graphicspath{{gfx/}}

\def\a{\hat{a}} 
\def\ad{\hat{a}^\dagger}
\def\bd{b^\dagger}

\def\n{\hat{n}} 

\def\B{\hat{B}} 

\def\D{\hat{D}} 

\renewcommand{\b}[1]{\mathbf{#1}} 
\newcommand{\bra}[1]{\langle #1 |}
\newcommand{\ket}[1]{| #1 \rangle}

\title{Quantum State Reduction by Matter-Phase-Related Measurements in
  Optical Lattices} 
\author[1,*]{Wojciech Kozlowski} 
\author[1,2]{Santiago F. Caballero-Benitez}
\author[1,3]{Igor B. Mekhov} 
\affil[1]{Department of Physics, Clarendon Laboratory, University
  of Oxford, Parks Road, Oxford OX1 3PU, United Kingdom}
\affil[2]{CONACYT, Instituto Nacional de Astrofísica, Óptica y
  Electrónica, Calle Luis Enrique Erro No. 1, Sta. María Tonantzintla,
  Pue. CP 72840, México}
\affil[3]{St. Petersburg State University, Universitetsky
  pr. 26,198504 St. Petersburg, Russia}

\affil[*]{wojciech.kozlowski@physics.ox.ac.uk}

\begin{abstract}
  A many-body atomic system coupled to quantized light is subject to
  weak measurement. Instead of coupling light to the on-site density,
  we consider the quantum backaction due to the measurement of
  matter-phase-related variables such as global phase coherence. We
  show how this unconventional approach opens up new opportunities to
  affect system evolution. We demonstrate how this can lead to a new
  class of final states different from those possible with dissipative
  state preparation or conventional projective measurements. These
  states are characterised by a combination of Hamiltonian and
  measurement properties thus extending the measurement postulate for
  the case of strong competition with the system's own evolution.
\end{abstract}
\begin{document}

\flushbottom
\maketitle

\thispagestyle{empty}

\section*{Introduction}

Ultracold gases trapped in optical lattices is a very successful and
interdisciplinary field of research \cite{lewenstein2007,
  bloch2008}. Whilst normally the atoms are manipulated using
classical light beams there is a growing body of work based on
coupling such systems to quantised optical fields exploring the
ultimate quantum level of light-matter coupling \cite{mekhov2012,
  ritsch2013}. This new regime of interactions has already led to a
host of fascinating phenomena, such as novel methods of
non-destructive probing of quantum states \cite{ eckert2008,
  roscilde2009, mekhov2009pra, dechiara2011, LP2012, hauke2013,
  rogers2014, elliott2015, kozlowski2015, atoms}, new quantum phases
and light-matter entanglement, \cite{moore1999, chen2009, baumann2010,
  wolke2012, schmidt2014, caballero2015, caballero2015njp,
  mazzucchi2015, caballero2016}, or an entirely new class of many-body
dynamics due to measurement backaction \cite{lee2014, blattmann2015,
  ashida2015, ashida2015a, ashida2016, mazzucchi2016, kozlowski2015a,
  mazzucchi2016a}. Furthermore, recent experimental breakthroughs in
coupling an optical lattice to a cavity demonstrate the significant
interest in studying this ultimate quantum regime of light-matter
interaction \cite{klinder2015, landig2015}.

Light scatters due to its interaction with the dipole moment of the
atoms which for off-resonant light results in an effective coupling
with atomic density, not the matter-wave amplitude. Therefore, it is
challenging to couple light to the phase of the matter-field, as is
typical in quantum optics for optical fields. Most of the existing
work on measurement couples directly to atomic density operators
\cite{LP2009, rogers2014, mekhov2012, ashida2015,
  ashida2015a}. However, it has been shown that it is possible to
couple to the the relative phase differences between sites in an
optical lattice by illuminating the bonds between them
\cite{kozlowski2015, caballero2015, caballero2015njp, mazzucchi2015,
  caballero2016, caballero2016a}. This is a multi-site generalisation
of previous double-well schemes \cite{cirac1996, castin1997,
  ruostekoski1997, ruostekoski1998, rist2012}, although the physical
mechanism is fundametally different as it involves direct coupling to
the interference terms caused by atoms tunnelling rather than
combining light scattered from different sources.

Coupling to phase observables in lattices has been proposed and
considered in the context of nondestructive probing and quantum
optical potentials. In this paper, we go beyond any previous work by
studying this new feature of optical lattice cavity systems in the
context of measurement backaction. The quantum trajectory approach to
backaction induced dynamics is not new in general and has attracted
significant experimental interest in single atom cavity
\cite{hood1998} and single qubit circuit \cite{murch2013, roch2014}
QED systems. However, its study in the context of many-body dynamics
is much more recent and has attracted significant theoretical interest
over the past years \cite{LP2010, LP2011, douglas2012, douglas2013,
  mekhov2009pra, mekhov2012, pedersen2014, lee2014, mazzucchi2016}.
Here, it is the novel combination of measurement backaction as the
physical mechanism driving the dynamics and phase coherence as the
observable, which the optical fields couple to, that provides a
completely new opportunity to affect and manipulate the quantum state.

In this paper we begin by presenting a simple quantum gas example. In
the second part we generalize our model and show a novel type of a
projection due to measurement which occurs even when there is
significant competition with the Hamiltonian dynamics. This projection
is fundamentally different to dissipative steady states, standard
formalism eigenspace projections or the quantum Zeno effect
\cite{misra1977, facchi2008, Raimond2010, Raimond2012, Signoles2014}
thus providing an extension of the measurement postulate to dynamical
systems subject to weak measurement. Such a measurement-based
preparation is unobtainable using the dissipative state engineering,
as the dissipation would completely destroy the coherence in this
case.

\section*{Results}

\subsection*{Quantum gas model}

We consider measurement of an ultracold gas of $N$ bosons trapped in
an optical lattice with period $a$ and $M$ sites \cite{mekhov2012}. We
focus on the one-dimensional case, but the general concept can be
easily applied to higher dimensions. The isolated system is described
by the Bose-Hubbard model with the Hamiltonian
\begin{equation}
  \hat{H}_0 = - J \sum_m \hat{p}_m + (U/2) \sum_m \n_m (\n_m - 1),
\end{equation}
where $\n_m = \bd_m b_m$ is the number operator at site $m$, $b_m$
annihilates an atom at site $m$,
$\hat{p}_m = \bd_m b_{m + 1} + b_m \bd_{m + 1}$, $J$ is the atom
hopping amplitude and $U$ the on-site interaction.

The atoms are illuminated with an off-resonant beam and light
scattered at a particular angle is selected and enhanced by a cavity
with decay rate $\kappa$ \cite{bux2013, kessler2014,
  landig2015a}. Just like in classical optics for light amplitude, the
Heisenberg annihilation operator of the scattered light is given by
$\a \sim \int u^*_\mathrm{out}(\b{r}) u_\mathrm{in} (\b{r}) \n(\b{r})
\mathrm{d} \b{r}$, where $\n(\b{r}) = \hat{\Psi}^\dagger(\b{r})
\hat{\Psi}(\b{r})$ is the atomic density operator, $\hat{\Psi}(\b{r})$
is the operator that annihilates a boson at $\b{r}$, and
$u_\mathrm{in,out}(\b{r})$ are the light mode functions for the
incoming and scattered beams. Expanding the matter-field operator in
terms of the Wannier functions of the lowest band, $\hat{\Psi} (\b{r})
= \sum_m b_m w(\b{r} - \b{r}_m)$, we can write $\a = C ( \D + \B )$
\cite{mekhov2009pra, mekhov2012}, where $C$ is the Rayleigh scattering
coefficient and
\begin{equation} 
  \D = \sum_m^K J_{m,m}\n_m, \quad \B = \sum_m^K J_{m,m
    + 1} \hat{p}_m,
\end{equation} 
the sum is over $K$ illuminated sites, and 
\begin{equation}
  J_{m,n}=\int w (\b{r}
  -\b{r}_m) u_{\mathrm{out}}^*(\b{r})u_{\mathrm{in}}(\b{r}) w (\b{r}
  -\b{r}_n) \, \mathrm{d} \b{r}.
\end{equation} 
We will consider the case when the quantum potential due to the cavity
light field is negligible (cavity detuning must be small compared to
$\kappa$ \cite{caballero2015}), but the photons leak from the cavity
and thus affect the system via measurement backaction instead
\cite{mekhov2012, mazzucchi2016}. This process can be modelled using a
quantum trajectory approach where each experimental run is simulated
using a stochastic Schr\"{o}dinger equation. Following the formalism
presented in Ref. \cite{mazzucchi2016} the system can be shown to
evolve according to $\hat{H} = \hat{H}_0 - i \kappa \ad \a$ and the
jump operator $\a$ is applied to the wave function whenever a photon
is detected. In a trajectory simulation the photodetection times are
determined using a Monte-Carlo method. Measurement backaction affects
the optical field which is entangled with the atoms and thus the
quantum gas is also affected, just like the particles in the
Einstein-Podolsky-Rosen thought experiment are affected by
measurements on its pair \cite{einstein1935}.

In general, it is easier for the light to couple to atom density that
is localised within the lattice rather than the density within the
bonds, i.e.~in between the lattice sites. This means that in most
cases $\D \gg \B$ and thus $\a \approx \D$. However, it is possible to
arrange the light geometry in such a way that scattering from the
atomic density operators within a lattice site is suppressed leading
to a situation where light is only scattered from these bonds leading
to an effective coupling to phase-related observables, $\a = C \B$
\cite{kozlowski2015}. This does not mean that light actually scatters
from the matter phase. Light scatters due to its interaction with the
dipole moment of the atoms which for off-resonant light and thus the
scattering is always proportional to the density
distribution. However, in an optical lattice, the interference of
matter waves between neighbouring sites leads to density modulations
which allows us to indirectly measure these phase observables. A brief
summary based on Ref. \cite{kozlowski2015} on how this is achieved is
available in the Supplementary Information online. Here, we will
summarise the results and focus on the effects of measurement
backaction due to such coupling.

\begin{figure}[htbp!]
  \centering
  \includegraphics[width=0.8\linewidth]{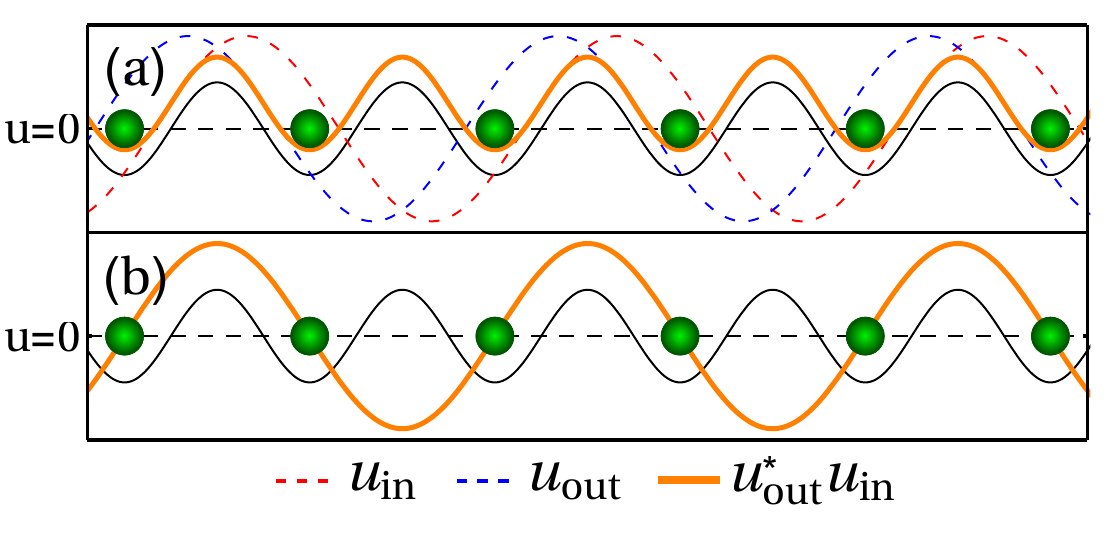}
  \caption{Light field arrangements which maximise coupling,
    $u_\mathrm{out}^*u_\mathrm{in}$, between lattice sites. The thin
    black line indicates the trapping potential (not to scale). (a)
    Arrangement for the uniform pattern $J_{m,m+1} = J_1$. (b)
    Arrangement for spatially varying pattern $J_{m,m+1}=(-1)^m J_2$;
    here $u_\mathrm{in}=1$ so it is not shown and $u_\mathrm{out}$ is
    real thus $u_\mathrm{out}^*u_\mathrm{in}=u_\mathrm{out}$.}
  \label{fig:setup}
\end{figure}

If we consider both incoming and outgoing beams to be standing waves,
$u_\mathrm{in,out} = \cos(k^x_\mathrm{in,out} x +
\varphi_\mathrm{in,out})$ we can suppress the $\D$-operator
contribution by crossing the beams at angles such that $x$-components
of the wavevectors are $k^x_\mathrm{in,out} = \pi/d$, and the phase
shifts satisfy $\varphi_\mathrm{in} + \varphi_\mathrm{out} = \pi$ and
$\varphi_\mathrm{in} - \varphi_\mathrm{out} = \arccos[\mathcal{F}
[w^2(\b{r})](2 \pi / a) / \mathcal{F} [w^2(\b{r})] (0) ]/2$, where
$\mathcal{F}[f(\b{r})]$ denotes a Fourier transform of $f(\b{r})$
\cite{kozlowski2015}. For clarity, this arrangement is illustrated in
Fig. \ref{fig:setup}(a). This ensures that $J_{m,m} = 0$ whilst
\begin{equation}
J_1 \equiv J_{m, m+1} = \mathcal{F} [w(\b{r} - \b{a}/2) w(\b{r} +
\b{a}/2)](2 \pi / a)/2, 
\end{equation}
a constant, and thus $\a = C \B_1$ ($\D = 0$, $\B = \B_1$) with
\begin{equation} 
  \B_1 = \sum_m^K J_1 \hat{p}_m = 2 J_1 \sum_k c^\dagger_k
  c_k \cos(ka),
\end{equation} 
where the second equality follows from converting to momentum space
via $b_m = \frac{1}{\sqrt{M}} \sum_k e^{-ikma} c_k$ and $c_k$
annihilates an atom with momentum $k$.

In order to correctly describe the dynamics of a single quantum
trajectory we have introduced a non-Hermitian term to the Hamiltonian,
$-i \kappa \ad \a$. As the jump operator itself, $\a$ is linearly
proportional to the atom density, the new term introduces a quadratic
atom density term on top of the nonlocality due to the global nature
of the probing. Therefore, in order to focus on the competition
between tunnelling and measurement backaction we do not consider the
other (standard) nonlinearity due to the atomic interactions: $U =
0$. Therefore, $\B_1$ is proportional to the Hamiltonian and both
operators have the same eigenstates, i.e.~Fock states in the momentum
basis. We can thus rewrite as
\begin{equation}
  \hat{H} = - \frac{J} {J_1} \B_1 - i \kappa |C|^2 \B_1^\dagger \B_1,
\end{equation}
which will naturally be diagonal in the $\B_1$ basis. Since it's
already diagonal we can easily solve its dynamics and show that the
probability distribution of finding the system in an eigenspace with
eigenvalue $B_1$ after $n$ photocounts at time $t$ is given by
\begin{equation} 
  p(B_1,n,t) = \frac{B_1^{2n}}{F(t)} \exp \left[ - 2
    \kappa |C|^2 B_1^2 t \right] p_0(B_1),
\end{equation} 
where $p_0(B_1)$ denotes the initial probability of observing $B_1$
\cite{mekhov2009pra, LP2010, LP2011} and $F(t)$ is the normalisation
factor. This distribution has peaks at $B_1 = \pm \sqrt{n/2\kappa
  |C|^2 t}$ and an initially broad distribution will narrow down
around these two peaks with time and successive photocounts. The final
state is in a superposition, because we measure the photon number,
$\ad \a$ and not field amplitude. Therefore, the measurement is
insensitive to the phase of $\a = C \B$ and we get a superposition of
$\pm B_1$. This means that the matter is still entangled with the
light as the two states scatter light with different phase which the
photocount detector cannot distinguish. However, this is easily
mitigated at the end of the experiment by switching off the probe beam
and allowing the cavity to empty out or by measuring the light phase
(quadrature) to isolate one of the components \cite{mekhov2009pra,
  mekhov2012, atoms}. Interestingly, this measurement will establish
phase coherence across the lattice, $\langle \bd_m b_n \rangle \ne 0$,
in contrast to density based measurements where the opposite is true,
Fock states with no coherences are favoured.

Unusually, we do not have to worry about the timing of the quantum
jumps, because the measurement operator commutes with the
Hamiltonian. This highlights an important feature of this measurement
- it does not compete with atomic tunnelling, and represents a quantum
nondemolition (QND) measurement of the phase-related observable
\cite{brune1992}. This is in contrast to conventional density based
measurements which squeeze the atom number in the illuminated region
and thus are in direct competition with the atom dynamics (which
spreads the atoms), thus requiring strong couplings for a projection
\cite{mazzucchi2016}. Here a projection is achieved at any measurement
strength which allows for a weaker probe and thus effectively less
heating and a longer experimental lifetime.


It is also possible to achieve a more complex spatial pattern of
$J_{m, m+1}$ \cite{kozlowski2015}. This way the observable will no
longer commute with the Hamiltonian (and thus will no longer be QND),
but will still couple to the phase related operators. This can be done
by tuning the angles such that the wavevectors are
$k^x_\mathrm{in} = 0$ and $k^x_\mathrm{out} = \pi/d$ and the phase
shift of the outgoing beam is $\varphi_\mathrm{out} = \pm \pi/d$.
This yields
\begin{equation}
  (-1)^m J_2 \equiv J_{m,m+1} = - (-1)^m \mathcal{F} [w(\b{r} - \b{a}/2) w(\b{r} +
  \b{a}/2)](\pi / a) \cos (\varphi_\mathrm{in}),
\end{equation}
where $J_2$ is a constant. Now $\a = C\B_2$ ($\D = 0$, $\B = \B_2$)
and the resulting coupling pattern is shown in
Fig. \ref{fig:setup}(b). The operator $\B_2$ is given by,
\begin{equation} 
  \B_2 = \sum_m^K (-1)^m J_2 \hat{p}_m = 2 i J_2 \sum_k c^\dagger_k
  c_{k - \pi/a} \sin(ka).
\end{equation} 
Note how the measurement operator now couples the momentum mode $k$
with the mode $k - \pi/a$. 

The measurement operator no longer commutes with the Hamiltonian so we
do not expect there to be a steady state as before. In order to
understand the measurement it will be easier to work in a basis in
which it is diagonal. We perform the transformation $\beta_k =
\frac{1}{\sqrt{2}} \left( c_k + i c_{k - \pi/a} \right)$,
$\tilde{\beta}_k = \frac{1}{\sqrt{2}} \left( c_k - i c_{k - \pi/a}
\right)$, which yields the following forms of the measurement operator
and the Hamiltonian: 
\begin{equation}
  \B_2 = 2 J_2 \sum_{\mathrm{RBZ}} \sin(ka) \left(
    \beta^\dagger_k \beta_k - \tilde{\beta}_k^\dagger \tilde{\beta}_k
  \right), 
\end{equation}
\begin{equation}
  \hat{H}_0 = 2 J \sum_{\mathrm{RBZ}} \cos(ka) \left(
    \beta_k^\dagger \tilde{\beta}_k + \tilde{\beta}^\dagger_k \beta_k
  \right),
\end{equation} 
where the summations are performed over the reduced Brilluoin Zone
(RBZ), $0 < k \le \pi/a$, to ensure the transformation is
canonical. We see that the measurement operator now consists of two
types of modes, $\beta_k$ and $\tilde{\beta_k}$, which are
superpositions of two momentum states, $k$ and $k - \pi/a$. Note how a
spatial pattern with a period of two sites leads to a basis with two
modes whilst a uniform pattern had only one mode, $c_k$.

\begin{figure}[htbp!]
  \centering
  \includegraphics[width=\linewidth]{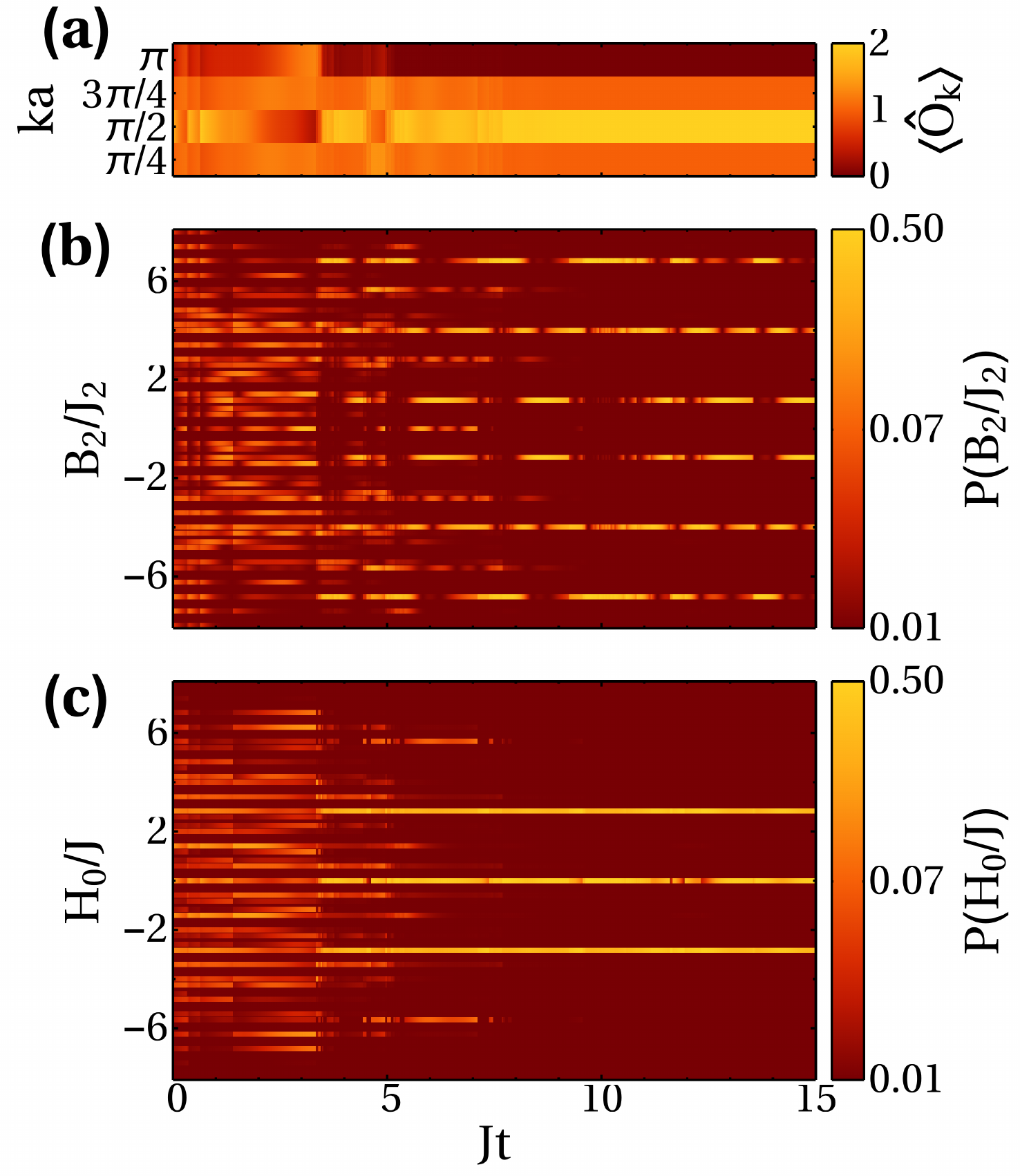}
  \caption{Subspace projections. Projection to a $\mathcal{P}_M$ space
    for four atoms on eight sites with periodic boundary
    conditions. The parameters used are $J=1$, $U=0$,
    $\kappa|C|^2=0.1$, and the initial state was
    $\ket{0,0,1,1,1,1,0,0}$. (a) The $\langle \hat{O}_k \rangle =
    \langle \n_k + \n_{k - \pi/a} \rangle$ distribution becomes fully
    confined to its subspace at $Jt \approx 8$ indicating the system
    has been projected. (b) Populations of the $\B_2$ eigenspaces. (c)
    Population of the $\hat{H}_0$ eigenspaces. Once the projection is
    achieved at $Jt\approx8$ we can see from (b-c) that the system is
    not in an eigenspace of either $\B_2$ or $\hat{H}_0$, but it
    becomes confined to some subspace. The system has been projected
    onto a subspace, but it is neither that of the measurement
    operator or the Hamiltonian.}
  \label{fig:projections}
\end{figure}

Trajectory simulations confirm that there is no steady state.
However, unexpectedly, for each trajectory we observe that the
dynamics always ends up confined to some subspace as seen in
Fig. \ref{fig:projections} which is not the same for each
trajectory. In general, this subspace is not an eigenspace of the
measurement operator or the Hamiltonian. In
Fig. \ref{fig:projections}(b) it in fact clearly consists of multiple
measurement eigenspaces. This clearly distinguishes it from the
typical projection formalism. It is also not the quantum Zeno effect
which predicts that strong measurement can confine the evolution of a
system as this subspace must be an eigenspace of the measurement
operator \cite{misra1977, facchi2008, Raimond2010, Raimond2012,
  Signoles2014}. Furthermore, the projection we see in
Fig. \ref{fig:projections} occurs for even weak measurement strengths
compared to the Hamiltonian's own evolution, a regime in which the
quantum Zeno effect does not happen. It is also possible to
dissipatively prepare quantum states in an eigenstate of a Hamiltonian
provided it is also a dark state of the jump operator, $\a | \Psi
\rangle = 0$, \cite{diehl2008}. However, this is also clearly not the
case here as the final state in Fig. \ref{fig:projections}(c) is not
only not confined to a single measurement operator eigenspace, it also
spans multiple Hamiltonian eigenspaces. Therefore, the dynamics
induced by $\a = C\B_2$ projects the system into some subspace, but
since this does not happen via any of the mechanisms described above
it is not immediately obvious what this subspace is. 

A crucial point is that whilst single quantum trajectories might not
have a steady state, for dissipative systems the density matrix will
in general have a steady state which can undergo phase transitions as
the dissipative parameters are varied \cite{kessler2012}. If we were
to average over many trajectories we would obtain such a steady state
for this system. However, we are concerned with measurement and not
dissipation. Whilst both are open systems, having knowledge of the
measurement outcome from the photodetector means we deal with pure
states that are the outcomes of individual measurements rather than an
ensemble average over all possible outcomes. This can reveal physical
effects which would be lost in a mixed state. The example in
Fig. \ref{fig:projections} shows how a single quantum trajectory can
become confined yet never approach any steady state - measurement and
tunnelling still compete, albeit in a limited subspace. This subspace
will not in general be the same for each experimental trajectory, but
once the subspace is chosen, the system will remain there. This is
analogous to a QND measurement in which a system after the first
projection will remain in its chosen eigenstate, but this eigenstate
is not determined until the first projection takes place. However, if
we were to look at the dissipative steady state (by averaging
expectation values over many quantum trajectories), we would not see
these subspaces at all, because the mixed state is an average over all
possible outcomes, and thus an average over all possible subspaces
which on a single trajectory level are mutually exclusive. Therefore,
here we will consider only individual experimental runs, which are not
steady states themselves, but rather the individual pure state
components of the dissipative steady state that are obtained via the
weak measurement of $\B_2$.

\subsection*{General model for the projection}

To understand this dynamics we will look at the master equation for
open systems described by the density matrix, $\hat{\rho}$,
\begin{equation}
  \label{eq:master}
  \dot{\hat{\rho}} = -i \left[\hat{H}_0, \hat{\rho} \right] + 2 \kappa
  \left[ \a \hat{\rho} \ad - \frac{1}{2} \left( \ad \a \hat{\rho} +
    \hat{\rho} \ad \a \right) \right],
\end{equation}
where $\a = C(\D + \B)$ as before. This equation describes the state
of the system if we discard all knowledge of the outcome which is
effectively an average over all possible stochastic quantum
trajectories. The commutator describes coherent dynamics due to the
isolated Hamiltonian and the remaining terms are due to
measurement. This is a convenient way to find features of the dynamics
common to every measurement trajectory.

We define the projectors of the measurement eigenspaces, $P_m$, which
have no effect on any of the (possibly degenerate) eigenstates of $\a$
with eigenvalue $a_m$, but annihilate everything else, thus $P_m =
\sum_{a_n = a_m} \ket{a_n} \bra{a_n},$ where $\ket{a_n}$ is an
eigenstate of $\a$ with eigenvalue $a_n$. Note that since $\a = C(\D +
\B)$ these projectors act on the matter state. This allows us to
decompose the master equation in terms of the measurement basis as a
series of equations $P_m \dot{\hat{\rho}} P_n$. For $m = n$, $P_m
\dot{\hat{\rho}} P_m = -i P_m \left[\hat{H}_0, \hat{\rho} \right]
P_m$, the measurement terms disappear which shows that a state in a
single eigenspace is unaffected by observation. On the other hand, for
$m \ne n$ the Hamiltonian evolution actively competes against
measurement. In general, if $\a$ does not commute with the Hamiltonian
then a projection to a single eigenspace $P_m$ is impossible.

We now define a new type of projector $\mathcal{P}_M = \sum_{m \in M}
P_m$, such that $\mathcal{P}_M \mathcal{P}_N = \delta_{M,N}
\mathcal{P}_M$ and $\sum_M \mathcal{P}_M = \hat{1}$ where $M$ denotes
some arbitrary subspace. The first equation implies that the subspaces
can be built from $P_m$ whilst the second and third equation specify
that these projectors do not overlap and that they cover the whole
Hilbert space. Furthermore, we will also require that $[\mathcal{P}_M,
  \hat{H}_0 ] = [\mathcal{P}_M, \a] = 0$. The second commutator simply
follows from the definition of $\mathcal{P}_M$, but the first one is
non-trivial. However, if we can show that $\mathcal{P}_M = \sum_{m \in
  M} \ket{h_m} \bra{h_m}$, where $\ket{h_m}$ is an eigenstate of
$\hat{H}_0$ then the commutator is guaranteed to be zero. Note that we
always have the trivial case where all these conditions are satisfied
and that is when there is only one such projector $\mathcal{P}_M =
\hat{1}$.

Assuming that it is possible to have non-trivial cases where
$\mathcal{P}_M \ne \hat{1}$ we can write the master equation as
\begin{equation} 
  \mathcal{P}_M \dot{\hat{\rho}} \mathcal{P}_N = -i \left[\hat{H}_0,
    \mathcal{P}_M \hat{\rho} \mathcal{P}_N \right] + 2 \kappa \left[
    \a \mathcal{P}_M \hat{\rho} \mathcal{P}_N \ad - \frac{1}{2} \left(
    \ad \a \mathcal{P}_M \hat{\rho} \mathcal{P}_N + \mathcal{P}_M
    \hat{\rho} \mathcal{P}_N \ad \a \right) \right].
\end{equation} 
Crucially, thanks to the commutation relations we were able to divide
the density matrix in such a way that each submatrix's time evolution
depends only on itself. When we partitioned the matrix with $P_m$ the
fact that the projectors did not commute with the operators meant that
we had terms of the form $P_m [\hat{H}_0, \hat{\rho}] P_n$ which
couple many different $P_m \hat{\rho} P_n$ submatrices with each
other.

We note that when $M = N$ the equations for $\mathcal{P}_M \hat{\rho}
\mathcal{P}_M$ will include subspaces unaffected by measurement,
i.e.~$P_m \hat{\rho} P_m$. Therefore, parts of the $\mathcal{P}_M
\hat{\rho} \mathcal{P}_M$ submatrices will also remain unaffected by
measurement. However, the submatrices $\mathcal{P}_M \hat{\rho}
\mathcal{P}_N$, for which $M \ne N$, are guaranteed to not contain
measurement-free subspaces thanks to the orthogonality of
$\mathcal{P}_M$. Therefore, for $M \ne N$ all elements of
$\mathcal{P}_M \hat{\rho} \mathcal{P}_N$ will experience a non-zero
measurement term whose effect is always
dissipative/lossy. Furthermore, these coherence submatrices
$\mathcal{P}_M \hat{\rho} \mathcal{P}_N$ are not coupled to any other
part of the density matrix and so they can never increase in
magnitude; the remaining coherent evolution is unable to counteract
the dissipative term without an `external pump' from other parts of
the density matrix. The combined effect is such that all
$\mathcal{P}_M \hat{\rho} \mathcal{P}_N$ for which $M \ne N$ will
always go to zero.

When all these cross-terms vanish, we are left with a density matrix
that is a mixed state of the form $\hat{\rho} = \sum_M \mathcal{P}_M
\hat{\rho} \mathcal{P}_M$. Since there are no coherences,
$\mathcal{P}_M \hat{\rho} \mathcal{P}_N$, this state contains only
classical uncertainty about which subspace, $\mathcal{P}_M$, is
occupied - there are no quantum superpositions between different
$\mathcal{P}_M$ spaces. Therefore, in a single measurement run we are
guaranteed to have a state that lies entirely within a subspace
defined by $\mathcal{P}_M$.

Before moving on to a specific example we will briefly discuss the
regime of validity of this result. In principle, this should be
applicable to any open system that can be described by the master
equation in Eq. \eqref{eq:master} as the projectors $P_m$ can be
constructed for any jump operator. The peculiar form of our operators,
namely that $\a = C (\D + \B)$, simply allows us to limit our system
to just the matter state, but is in general not necessary to obtain
the result above. In fact, QND measurements, such as the one seen in
the previous section, are another special case where each of the new
projectors $\mathcal{P}_M$ is made of a sum of projectors $P_m$ in a
single degenerate subspace. Therefore, the existence of these emergent
subspaces relies on exactly the same physical approximations as the
master equation and is simply one of the properties of Markovian open
systems. However, the existence of these trivial cases alone does not
justify the introduction of a new set of projectors. Furthermore, the
derivation alone does not help us in identifying what systems might
have non-trivial subspaces or whether any even exist. Since this
result applies to any system described by a master equation which will
always exhibit the trivial cases of the identity and QND measurement
projectors, it is unclear whether it is in general possible to predict
which Hamiltonians might have these non-trivial emergent subspaces.

However, it turns out that such a non-trivial case is indeed possible
for our $\hat{H}_0$ and $\a = C\B_2$ and we can see the effect in
Fig. \ref{fig:projections}. Whilst the result is general and
applicable to any Markovian system, we identified the first
non-trivial case only for phase observable measurements in an optical
lattice. This is thanks to the fact that the
measurement operator is similar in form to the Hamiltonian, but at the
same time it does not commute with it (otherwise we would have a QND
measurement).

In Fig. \ref{fig:projections} we can clearly see how a state that was
initially a superposition of a large number of eigenstates of both
operators becomes confined to some small subspace that is neither an
eigenspace of $\a$ or $\hat{H}_0$. In this case the projective spaces,
$\mathcal{P}_M$, are defined by the parities (odd or even) of the
combined number of atoms in the $\beta_k$ and $\tilde{\beta}_k$ modes
for different momenta $0 < k < \pi/a$ that are distinguishable to
$\B_2$. The explanation requires careful consideration of where the
eigenstates of the two operators overlap and is described in Section
S3 of the Supplementary Information online.

\begin{figure}[htbp!]
  \centering
  \includegraphics[width=0.8\linewidth]{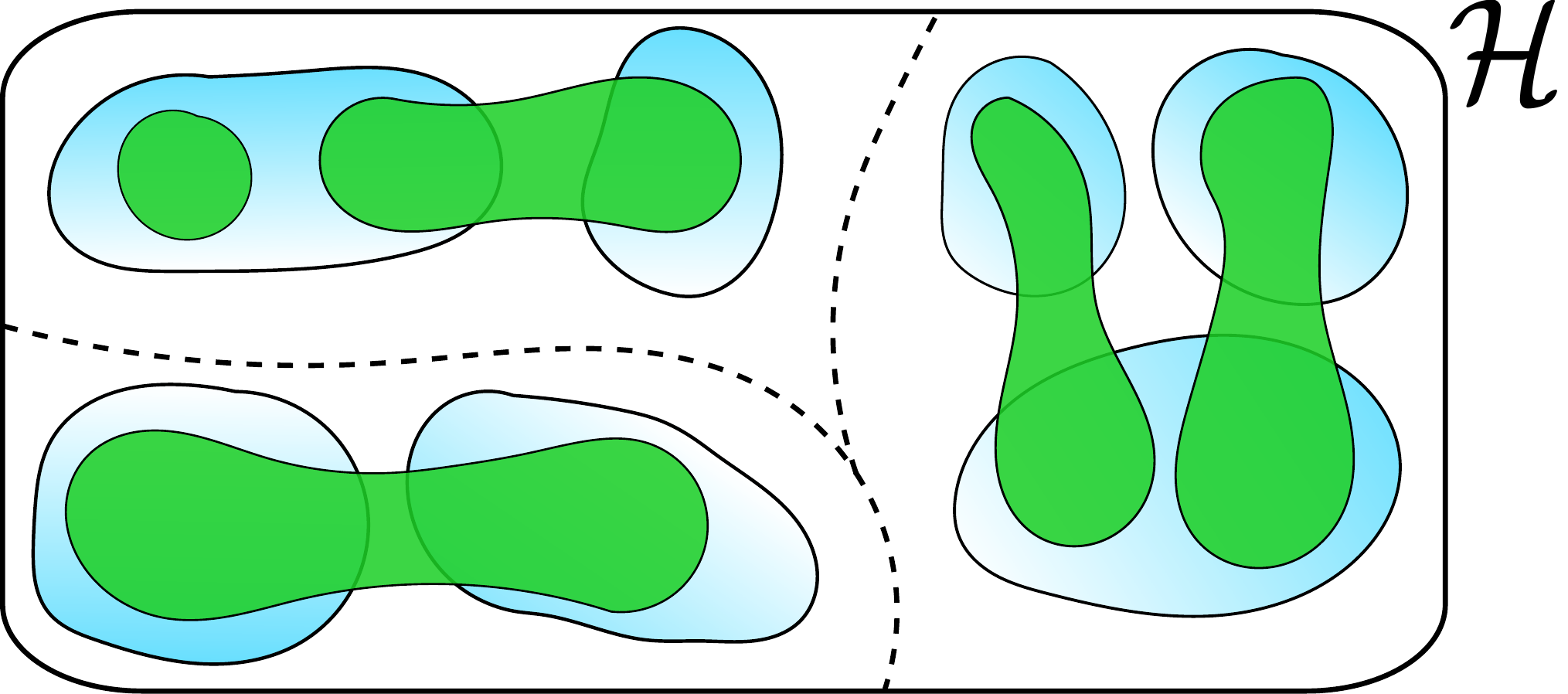}
  \caption{A visual representation of the projection spaces of the
    measurement. The light blue areas (bottom layer) are $R_m$, the
    eigenspaces of $\hat{O}$. The green areas are measurement
    eigenspaces, $P_m$, and they overlap non-trivially with the $R_m$
    subspaces. The $\mathcal{P}_M$ projection space boundary (dashed
    line) runs through the Hilbert space, $\mathcal{H}$, where there
    is no overlap between $P_m$ and $R_m$.}
   \label{fig:spaces}
\end{figure}

To understand the physical meaning of these projections we define an
operator $\hat{O}$ with eigenspace projectors $R_m$, which commutes
with both $\hat{H}_0$ and $\a$. Physically this means that $\hat{O}$
is a compatible observable with $\a$ and corresponds to a quantity
conserved by the Hamiltonian. The fact that $\hat{O}$ commutes with
the Hamiltonian implies that the projectors can be written as a sum of
Hamiltonian eigenstates $R_m = \sum_{h_i = h_m} \ket{h_i} \bra{h_i}$
and thus a projector $\mathcal{P}_M = \sum_{m \in M} R_m$ is
guaranteed to commute with the Hamiltonian and similarly since
$[\hat{O}, \a] = 0$ $\mathcal{P}_M$ will also commute with $\a$ as
required. Therefore, $\mathcal{P}_M = \sum_{m \in M} R_m = \sum_{m \in
  M} P_m$ will satisfy all the necessary prerequisites. This is
illustrated in Fig. \ref{fig:spaces}.

In the simplest case the projectors $\mathcal{P}_M$ can consist of
only single eigenspaces of $\hat{O}$, $\mathcal{P}_M = R_m$. The
interpretation is straightforward - measurement projects the system
onto a eigenspace of an observable $\hat{O}$ which is a compatible
observable with $\a$ and corresponds to a quantity conserved by the
coherent Hamiltonian evolution. However, this may not be possible and
we have the more general case when $\mathcal{P}_M = \sum_{m \in M}
R_m$. In this case, one can simply think of all $R_{m \in M}$ as
degenerate just like eigenstates of the measurement operator, $\a$,
that are degenerate, can form a single eigenspace $P_m$. However,
these subspaces will correspond to different eigenvalues of $\hat{O}$
distinguishing it from conventional projections.

In our case, it is apparent from the form of $\B_2$ and $\hat{H}_0$
that
$\hat{O}_k = \beta_k^\dagger \beta_k + \tilde{\beta}_k^\dagger
\tilde{\beta_k} = \n_k + \n_{k - \pi/a}$ commutes with both operators
for all $k$. Thus, we can easily construct
$\hat{O} = \sum_\mathrm{RBZ} g_k \hat{O}_k$ for any arbitrary
$g_k$. Its eigenspaces, $R_m$, can then be easily constructed and
their relationship with $P_m$ and $\mathcal{P}_M$ is illustrated in
Fig. \ref{fig:spaces} whilst the time evolution of
$\langle \hat{O}_k \rangle$ for a sample trajectory is shown in
Fig. \ref{fig:projections}(a). These eigenspaces are composed of Fock
states in momentum space that have the same number of atoms within
each pair of $k$ and $k - \pi/a$ modes. The projectors $\mathcal{P}_M$
consist of many such eigenspaces leading to the case where we can only
distinguish between the spaces that have different parities of
$\hat{O}_k$.

\section*{Experimental considerations}

Before concluding this paper, it is worthwhile to consider the
experimental difficulties in realising such an experiment. First, we
note that there are two recent experiments that have successfully
obtained an ultracold gas in an optical lattice coupled to a high-Q
cavity \cite{klinder2015, landig2015}. The main major concern is
photon detector inefficiency. It has been shown \cite{mazzucchi2016a}
that as long as there is a sufficient number of photons detected such
that the true instantaneous rate can be reliably estimated it is
possible to use detectors with very low efficiencies. Another,
possible issue is the sensitivity of the relative angle between the
cavity and the probe beams. Generally, the most interesting
arrangements, such as the two cases used in this paper, correspond to
easily identifiable scattering features such as diffraction maxima and
minima, and thus they should be easy to identify and tune. However, it
is also possible to obtain identical jump operators with a homodyne
detection scheme in which instead of angles, one has to tune the local
oscillator phase which might potentially be easier to fine tune in an
experiment \cite{kozlowski2015}. Finally, one might also be concerned
with possible dephasing due to scattering outside of the
cavity. However, cavities used by experiments such as those in
Ref. \cite{baumann2010, landig2015} have a Purcell factor of $\sim$100
and probe-atom detunings in the MHz range. Thus, any scattering
outside of the cavity can be safely neglected \cite{baumann2010}.

\section*{Discussion}

In summary we have investigated measurement backaction resulting from
coupling light to an ultracold gas's phase-related observables. We
demonstrated how this can be used to prepare the Hamiltonian
eigenstates even if significant tunnelling is occuring as the
measurement can be engineered to not compete with the system's
dynamics. Furthermore, we have shown that when the observable of the
phase-related quantities does not commute with the Hamiltonian we
still project to a specific subspace of the system that is neither an
eigenspace of the Hamiltonian or the measurement operator. This is in
contrast to quantum Zeno dynamics \cite{misra1977, facchi2008,
  Raimond2010, Raimond2012, Signoles2014} or dissipative state
preparation \cite{diehl2008}. We showed that this projection is
essentially an extension of the measurement postulate to weak
measurement on dynamical systems where the competition between the two
processes is significant.

\section*{Supplementary Information}

\renewcommand\theequation{S\arabic{equation}}
\setcounter{equation}{0}

\renewcommand\thefigure{S\arabic{figure}}
\setcounter{figure}{0}

\renewcommand\thetable{S\arabic{table}}
\setcounter{table}{0}

\renewcommand\thesection{S\arabic{section}}
\setcounter{section}{0}

\section{Suppressing the effective coupling to atomic density}

In the main text we showed that $\hat{a} = C ( \hat{D} + \hat{B} )$,
where
\begin{equation}
  C = \frac{g_\mathrm{out} g_\mathrm{in} a_0}{ \Delta_a \left(
    \Delta_p + i \kappa \right)},
\end{equation}
$g_\mathrm{out,in}$ are the atom-light coupling constants for the
outgoing and incoming beams, $\Delta_a$ is the detuning between the
incoming probe beam and the atomic resonance frequency, $\Delta_p$ is
the detuning between the incoming probe beam and the outgoing cavity
beam, $a_0$ is the amplitude of the coherent probe beam, and $\kappa$
is the cavity decay rate. However, we are only interested in the case
when $\hat{a} = C\hat{B}$. Therefore, we need to find the conditions
under which this is true. For clarity we will consider a 1D lattice,
but the results can be applied and generalised to higher
dimensions. Central to engineering the $\hat{a}$ operator are the
coefficients $J_{m,n}$ given by
\begin{equation}
  \label{Jcoeff}
  J_{m,n}=\int w (\b{r} -\b{r}_m)
  u_{\mathrm{out}}^*(\b{r})u_{\mathrm{in}}(\b{r}) w (\b{r} -\b{r}_n) \,
  \mathrm{d} \b{r},
\end{equation}
where $w(\b{r})$ are the Wannier functions of the lowest band,
$u_\mathrm{in,out} (\b{r})$ are the light mode functions of the
incoming and outgoing beams, and $\b{r}$ is the position vector. The
operators $\hat{B}$ and $\hat{D}$ depend on the values of $J_{m,m+1}$
and $J_{m,m}$ respectively and are given by
\begin{equation} 
  \D = \sum_m^K J_{m,m}\n_m,
\end{equation} 
\begin{equation}
  \B = \sum_m^K J_{m,m + 1} \left( \bd_m b_{m+1} + b_m \bd_{m+1} \right),
\end{equation} 
where $b_m$ annihilates an atom at site $m$, and $\n_m = \bd_m b_m$ is
the number operator at site $m$. These $J_{m,n}$ coefficients are
determined by the convolution of the light mode product,
$u_\mathrm{out}^*({\bf r})u_\mathrm{in}({\bf r})$ with the relevant
Wannier function overlap $w(\b{r} - \b{r}_m) w(\b{r} - \b{r}_n)$. For
the $\hat{B}$ operator we calculate the convolution with the nearest
neighbour overlap, $W_1({\bf r}) \equiv w({\bf r} - {\bf a}/2) w({\bf
  r}+{\bf a}/2)$, where $\b{a}$ is the site separation vector, and for
the $\hat{D}$ operator we calculate the convolution with the square of
the Wannier function at a single site, $W_0({\bf r}) \equiv w^2({\bf
  r})$. Therefore, in order to enhance the $\hat{B}$ term we need to
maximise the overlap between the light modes and the nearest neighbour
Wannier overlap, $W_1({\bf r})$. This can be achieved by concentrating
the light between the sites rather than at atom positions.

In order to calculate the $J_{m,n}$ coefficients it is necessary to
perform numerical calculations using realistic Wannier
functions. However, it is possible to gain some analytic insight into
the behaviour of these values by looking at the Fourier transforms of
the Wannier function overlaps, $\mathcal{F}[W_{0,1}]({\bf k})$. This
is because the light mode product, $u_\mathrm{out}^*({\bf
  r})u_\mathrm{in}({\bf r})$, can be in general decomposed into a sum
of oscillating exponentials of the form $e^{i {\bf k} \cdot {\bf r}}$
making the integral in Eq. (\ref{Jcoeff}) a sum of Fourier transforms
of $W_{0,1}({\bf r})$.

\begin{figure}[htbp!]
  \centering
  \includegraphics[width=0.8\linewidth]{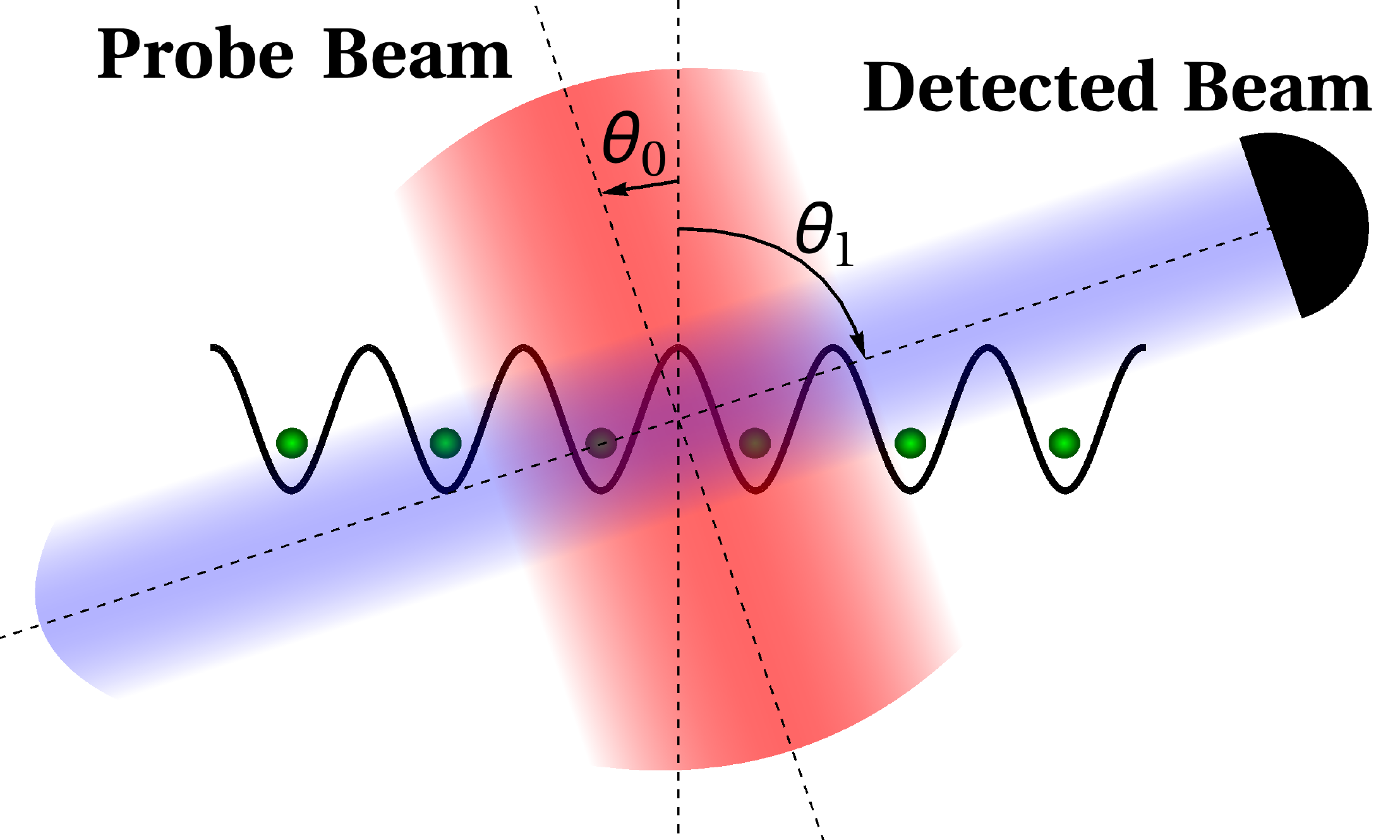}
  \caption{Setup. Atoms in an optical lattice are illuminated by a
    probe beam. The light scatters in free space or into a cavity and
    is measured by a detector.}
  \label{fig:Setup}
\end{figure}

We consider a setup shown in Fig. \ref{fig:Setup} and take both the
detected and probe beam to be standing waves,
$u_\mathrm{in,out}(\b{r}) = \cos(\b{k}_\mathrm{in,out} \cdot \b{r} +
\varphi_\mathrm{in,out})$, where $\b{k}$ is the wavevector of the beam
and $\varphi$ is a constant phase shift. This gives the following
expressions for the $\hat{D}$ and $\hat{B}$ operators
\begin{equation}
\label{FTD}
  \hat{D} = \frac{1}{2}[\mathcal{F}[W_0](k_-)\sum_m\hat{n}_m\cos(k_-
    x_m +\varphi_-) +\mathcal{F}[W_0](k_+)\sum_m\hat{n}_m\cos(k_+ x_m
    +\varphi_+)],
\end{equation}
\begin{equation}
  \label{FTB}
  \hat{B} = \frac{1}{2}[\mathcal{F}[W_1](k_-)\sum_m\hat{p}_m\cos(k_-
    x_m +\frac{k_-a}{2}+\varphi_-)
    +\mathcal{F}[W_1](k_+)\sum_m\hat{p}_m\cos(k_+ x_m
    +\frac{k_+a}{2}+\varphi_+)],
\end{equation}
where $x_m = ma$, $k_\pm = k_{\mathrm{in},x} \pm k_{\mathrm{out},x}$,
$k_{(\mathrm{in},\mathrm{out})x} = k_{\mathrm{in},\mathrm{out}}
\sin(\theta_{\mathrm{in},\mathrm{out}})$,
$\hat{p}_m=b^\dag_mb_{m+1}+b_mb^\dag_{m+1}$, and
$\varphi_\pm=\varphi_\mathrm{in} \pm \varphi_\mathrm{out}$. The key
result is that the $\hat{B}$ operator is phase shifted by $k_\pm d/2$
with respect to the $\hat{D}$ operator since it depends on the
amplitude of light in between the lattice sites and not at the
positions of the atoms, allowing to decouple them at specific angles.

Firstly, we will use this result to show how one can obtain the
uniform pattern for which $\B = \B_1$, where 
\begin{equation} 
  \B_1 = \sum_m^K J_1 \left( \bd_m b_{m+1} + b_m \bd_{m+1} \right),
\end{equation} 
i.e.~$J_{m,m+1} = J_1$ This can be achieved by crossing the light
modes such that $\theta_\mathrm{in} = -\theta_\mathrm{out}$ and
$k_{\mathrm{in},x} = k_{\mathrm{out},x} = \pi/a$ and choosing the
light mode phases such that $\varphi_+ = \pi$. In order to make the
$\hat{B}$ contribution to light scattering dominant we need to set
$\hat{D} = 0$ which from Eq. \eqref{FTD} we see is possible if
$\varphi_- =
\arccos[\mathcal{F}[W_0](2\pi/a)/\mathcal{F}[W_0](0)]/2$. This
arrangement of light modes maximizes the interference signal,
$\hat{B}$, by suppressing the density signal, $\hat{D}$, via
interference compensating for the spreading of the Wannier functions
and leads to the parameter value $J_1 =
\mathcal{F}[W_1](2\pi/a)/2$. The light mode patterns are illustrated
in the main text in Fig. 1(a).

Secondly, we show that we can have a spatially varying pattern for
which $\B  = \B_2$, where
\begin{equation} 
  \B_2 = \sum_m^K (-1)^m J_2 \left( \bd_m b_{m+1} + b_m \bd_{m+1} \right).
\end{equation} 
We consider an arrangement where the beams are arranged such that
$k_{\mathrm{in},x} = 0$ and $k_{\mathrm{out},x} = \pi/a$ which gives
the following expressions for the density and interference terms
\begin{eqnarray}
  \label{DMin}
  \hat{D} = \mathcal{F}[W_0](\pi/a) \sum_m (-1)^m \hat{n}_m
  \cos(\varphi_\mathrm{in})  \cos(\varphi_\mathrm{out}) \nonumber \\
  \hat{B} = -\mathcal{F}[W_1](\pi/a) \sum_m (-1)^m \hat{p}_m 
  \cos(\varphi_\mathrm{in}) \sin(\varphi_\mathrm{out}).
\end{eqnarray}
It is clear that for $\varphi_\mathrm{out} = \pm \pi/2$, $\hat{D} =
0$, which is intuitive as this places the lattice sites at the nodes
of the mode $u_\mathrm{out}({\bf r})$ and yields the parameter value
$J_2 = -\mathcal{F}[W_1](\pi/a) \cos(\varphi_\mathrm{in})$. This is a
diffraction minimum as the light amplitude is zero, $\langle \hat{B}
\rangle = 0$, because contributions from alternating inter-site
regions interfere destructively. However, the intensity $\langle \ad
\a \rangle = |C|^2 \langle \hat{B}^2 \rangle$ is proportional to the
variance of $\hat{B}$ and is non-zero. The light mode patterns are
illustrated in the main text in Fig. 1(b).

\section{Finding the measurement projection subspaces}

The main text defines the projectors $\mathcal{P}_M = \sum_{m \in M}
P_m$, where $P_m$ are the projectors onto the $\a$ eigenspaces, such
that $\sum_M \mathcal{P}_M = \hat{1}$, $\mathcal{P}_M \mathcal{P}_N =
\delta_{M,N} \mathcal{P}_M$, $[\mathcal{P}_M, \hat{H}_0] = 0$, and
$[\mathcal{P}_M, \a] = 0$. To find $\mathcal{P}_M$ we need to identify
the subspaces $M$ which satisfy the following relation $\sum_{m \in M}
P_m = \sum_{m \in M} \ket{h_m} \bra{h_m}$, where $\ket{h_m}$ are the
eigenstates of $\hat{H}_0$. This can be done iteratively by (i)
selecting some $P_m$, (ii) identifying the $\ket{h_m}$ which overlap
with this subspace, (iii) identifying any other $P_m$ which also
overlap with these $\ket{h_m}$ from step (ii). We repeat (ii)-(iii)
for all the $P_m$ found in (iii) until we have identified all the
subspaces $P_m$ linked in this way and they will form one of our
$\mathcal{P}_M$ projectors. If $\mathcal{P}_M \ne 1$ then there will
be other subspaces $P_m$ which we have not included so far and thus we
repeat this procedure on the unused projectors until we identify all
$\mathcal{P}_M$. Computationally this can be straightforwardly solved
with some basic algorithm that can compute the connected components of
a graph.

The above procedure, whilst mathematically correct and always
guarantees to generate the projectors $\mathcal{P}_M$, is very
unintuitive and gives poor insight into the nature or physical meaning
of $\mathcal{P}_M$. In order to get a better understanding of these
subspaces we will use another result from the main text. We showed
that for an operator $\hat{O}$ with eigenspace projectors $R_m$ for
which $[\hat{O}, \hat{H}_0] = 0$, and $[\hat{O}, \a] = 0$, then we can
write the subspace projectors as $\mathcal{P}_M = \sum_{m \in M} R_m =
\sum_{m \in M} P_m$.

We are interested in identifying these subspaces for the operator
$\B_2$ given by
\begin{align} 
  \B_2 = & \sum_m^K (-1)^m J_2 \left( \bd_m b_{m+1} + b_m \bd_{m+1}
  \right) \nonumber \\ = & 2 i J_2 \sum_k c^\dagger_k c_{k - \pi/a}
  \sin(ka).
\end{align} 
We have identified that for $\B_2$, an operator $\hat{O}$ that
commutes with both the measurement operator and the Hamiltonian is
given by $\hat{O} = \sum_\mathrm{RBZ} g_k \hat{O}_k$, where $\hat{O}_k
= \n_k + \n_{k - \pi/a}$, for any arbitrary constants $g_k$. The
subspaces $R_m$ of this operator simply consist of momentum space Fock
states that have the same number of atoms in each ($k$, $k - \pi/a$)
pair of momenta. However, it turns out that the $\mathcal{P}_M$
consist of multiple such subspaces complicating the picture.

Firstly, since $\B_2$ contains $\sin(ka)$ coefficients atoms in
different $k$ modes that have the same $\sin(ka)$ value are
indistinguishable to the measurement and will lie in the same $P_m$
eigenspaces. This will happen for the pairs ($k$, $\pi/a -
k$). Therefore, the $R_m$ spaces that have the same
$\hat{O}_k + \hat{O}_{\pi/a - k}$ eigenvalues must belong to the same
$\mathcal{P}_M$.

Secondly, if we re-write these operators in terms of the $\beta_k$ and
$\tilde{\beta}_k$ modes we get
\begin{equation} 
  \B_2 = 2 J_2 \sum_{\mathrm{RBZ}} \sin(ka) \left(
    \beta^\dagger_k \beta_k - \tilde{\beta}_k^\dagger \tilde{\beta}_k
  \right),
\end{equation}
\begin{equation} 
  \hat{O} = \sum_{\mathrm{RBZ}} g_k \left(
    \beta^\dagger_k \beta_k + \tilde{\beta}_k^\dagger \tilde{\beta}_k
  \right),
\end{equation}
and so it's not hard to see that $\B_{2,k} = (\beta^\dagger_k \beta_k
- \tilde{\beta}_k^\dagger \tilde{\beta}_k)$ will have the same
eigenvalues for different values of $\hat{O}_k = \beta^\dagger_k
\beta_k + \tilde{\beta}_k^\dagger \tilde{\beta}_k$. Specifically, if a
given subspace $R_m$ corresponds to the eigenvalue $O_k$ of
$\hat{O}_k$ then the possible values of $B_{2,k}$ will be $\{-O_k,
-O_k + 2, ..., O_k - 2, O_k\}$. Thus, we can see that all $R_m$ with
even values of $O_k$ will share $B_{2,k}$ eigenvalues and thus they
will overlap with the same $P_m$ subspaces. The same is true for odd
values of $O_k$. However, $R_m$ with an even value of $O_k$ will never
have the same value of $B_{2,k}$ as a subspace with an odd value of
$O_k$. Therefore, a single $\mathcal{P}_M$ will contain all $R_m$ that
have the same parities of $O_k$ for all $k$, e.g.~if it includes the
$R_m$ with $O_k = 6$, it will also include the $R_m$ for which $O_k =
0, 2, 4, 6, ..., N$, where $N$ is the total number of atoms.

Finally, the $k = \pi/a$ mode is special, because $\sin(\pi) = 0$
which means that $B_{2,k=\pi/a} = 0$ always. This in turn implies that
all possible values of $O_{\pi/a}$ are degenerate to the
measurement. Therefore, we exclude this mode when matching the
parities of the other modes.

To illustrate the above let us consider a specific example. Let us
consider two atoms, $N=2$, on eight sites $M=8$. This configuration
has eight momentum modes $ka = \{-\frac{3\pi}{4}, -\frac{\pi}{2},
-\frac{\pi}{4}, 0, \frac{\pi}{4}, \frac{\pi}{2}, \frac{3\pi}{4},
\pi\}$ and so the RBZ has only four modes $\mathrm{RBZ} :=
\{\frac{\pi}{4}, \frac{\pi}{2}, \frac{3\pi}{4}, \pi\}$. There are 10
different ways of splitting two atoms into these four modes and thus
we have 10 different $R_m = \{O_{\pi/4a}, O_{\pi/2a} ,O_{3\pi/4a}
,O_{\pi/a}\}$ eigenspaces of $\hat{O}$ and they are shown in Table
\ref{tab:Rm}. In the third column we have also listed the eigenvalues
of the $\B_2$ eigenstates that lie within the given $R_m$.

\begin{table}[htbp!]
  \centering
  \begin{tabular}{l c c}
    \toprule
    $m$ & $R_m$ & Possible values of $B_\mathrm{min}$ \\ \midrule
    0 & $\{2,0,0,0\}$ & $ -\sqrt{2}, 0, \sqrt{2}  $ \\
    1 & $\{1,1,0,0\}$ & $  -\frac{1 + \sqrt{2}}{\sqrt{2}}, -\frac{1
                        - \sqrt{2}}{\sqrt{2}}, \frac{1
                        - \sqrt{2}}{\sqrt{2}}, \frac{1 +
                        \sqrt{2}}{\sqrt{2}}  $ \\ 
    2 & $\{1,0,1,0\}$ & $  -\sqrt{2}, 0, \sqrt{2}  $ \\
    3 & $\{1,0,0,1\}$ & $  -\frac{1}{\sqrt{2}},
                        \frac{1}{\sqrt{2}}  $ \\
    4 & $\{0,2,0,0\}$ & $  -2, 0, 2  $ \\
    5 & $\{0,1,1,0\}$ & $  -\frac{1 + \sqrt{2}}{\sqrt{2}}, -\frac{1
                        - \sqrt{2}}{\sqrt{2}}, \frac{1
                        - \sqrt{2}}{\sqrt{2}}, \frac{1 +
                        \sqrt{2}}{\sqrt{2}} $ \\
    6 & $\{0,1,0,1\}$ & $  -1, 1  $ \\
    7 & $\{0,0,2,0\}$ & $  -\sqrt{2}, 0, \sqrt{2}  $ \\
    8 & $\{0,0,1,1\}$ & $  -\frac{1}{\sqrt{2}},
                        \frac{1}{\sqrt{2}}  $ \\
    9 & $\{0,0,0,2\}$ & $0$ \\
    \bottomrule
  \end{tabular}
  \caption{A list of all $R_m$ eigenspaces for $N = 2$ atoms at $M =
    8$ sites. The third column displays the eigenvalues of all the
    eigenstates of $\B_2$ that lie in the given $R_m$.}
  \label{tab:Rm}
\end{table}

We note that $ka = \pi/4$ will be degenerate with $ka = 3\pi/4$ since
$\sin(ka)$ is the same for both. Therefore, we already know that we
can combine $(R_0, R_2, R_7)$, $(R_1, R_5)$, and $(R_3, R_8)$, because
those combinations have the same $O_{\pi/4a} + O_{3\pi/4a}$
values. This is very clear in the table as these subspaces span
exactly the same values of $B_2$.

Now we have to match the parities. Subspaces that have the same parity
combination for the pair $(O_{\pi/4a} + O_{3\pi/4a}, O_{\pi/2a})$ will
be degenerate in $\mathcal{P}_M$. Note that we excluded $O_{\pi/a}$,
because as we discussed earlier they are all degenerate due to
$\sin(\pi) = 0$. Therefore, the (even,even) subspace will include
$(R_0, R_2, R_4, R_7, R_9)$, the (odd,even) will contain $(R_3, R_8)$,
the (even, odd) will contain $(R_6)$ only, and the (odd, odd) contains
$(R_1, R_5)$. These overlaps should be evident from the table as we
can see that these combinations combine all $R_m$ that contain any
eigenstates of $\B_2$ with the same eigenvalues.

Therefore, we have end up with four distinct $\mathcal{P}_M$ subspaces
\begin{align}
\mathcal{P}_\mathrm{even,even} = & R_0 + R_2 + R_4 + R_7 + R_9
  \nonumber \\
\mathcal{P}_\mathrm{odd,even} = & R_3 + R_8 \nonumber \\
\mathcal{P}_\mathrm{even,odd} = & R_6 \nonumber \\
\mathcal{P}_\mathrm{odd,odd} = & R_1 + R_5 \nonumber.
\end{align}
At this point it should be clear that these projectors satisfy all our
requirement. The conditions $\sum_M \mathcal{P}_M = 1$ and
$\mathcal{P}_M \mathcal{P}_N = \delta_{M,N} \mathcal{P}_M$ should be
evident from the form above. The commutator requirements are also
easily satisfied since the subspaces $R_m$ are of an operator that
commutes with both the Hamiltonian and the measurement operator. And
finally, one can also verify using the table that all of these
projectors are built from complete subspaces of $\B_2$ (i.e.~each
subspace $P_m$ belongs to only one $\mathcal{P}_M$) and thus
$\mathcal{P}_M = \sum_{m \in M} P_m$.


\section*{Acknowledgements}

We are grateful to D. A. Ivanov for constructive feedback on the
manuscript. The authors are grateful to EPSRC (DTA and
EP/I004394/1). S.F.C.-B acknowledges support from Cátedras CONACYT
para Jóvenes Investigadores project No. 551.

\section*{Author contributions statement}

W.K. is the lead author and performed the analysis and numerical
simulations. S.F.C.-B. and I.B.M. supervised the work. All authors
generated ideas for this paper and discussed the text at all stages.

\section*{Competing financial interests}

The authors declare no competing financial interests.

\end{document}